\documentclass[12pt]{article}
\newcommand{\bee}{\begin{equation}}                                             
\newcommand{\ee}{\end{equation}}                                                
\newcommand{\ba}{\begin{array}}                                                 
\newcommand{\ea}{\end{array}}                                                   
\newcommand{\bea}{\begin{eqnarray}}                                             
\newcommand{\eea}{\end{eqnarray}}

\begin{document}
\thispagestyle{empty}
\title{
Questionable Arguments for the Correctness of Perturbation Theory
in Non-Abelian Models
}                                                                               
\author{Adrian Patrascioiu \\
Physics Department, University of Arizona \\
Tucson, AZ 85721, U.S.A.\\
and\\
Erhard Seiler \\
Max-Planck-Institut f\"ur Physik (Werner-Heisenberg-Institut)\\
F\"ohringer Ring 6, 80805 Munich, Germany}
\maketitle                                                                      
{                                                                               

\begin{abstract}
\noindent
We analyze the arguments put forward recently by Niedermayer et al 
in favor of the correctness of conventional perturbation theory in 
non-Abelian models and supposedly showing that our super-instanton 
counterexample was sick. We point out that within
their own set of assumptions, the proof of Niedermayer et al regarding 
the correctness of perturbation theory is incorrect and provide a 
correct proof under more restrictive assumptions. We reply also to 
their claim that the S-matrix bootstrap approach of Balog et al 
supports the existence of asymptotic freedom in the O(3) model.
\end{abstract}
\newpage

A recent paper by Niedermayer, Niedermaier and Weisz entitled
"Questionable and Unquestionable in the Perturbation Theory of 
Non-Abelian Models" \cite{nnw} purports to show that our criticism of the 
standard dogma regarding the alleged difference between Abelian and 
non-Abelian models is exaggerated and that there are good reasons to
believe the orthodoxy. It is a positive development that members of the
high energy physics community are now beginning to pay attention to the 
fact that this central issue for particle physics remains  
mathematically unresolved and that at least some arguments are needed 
in support of the conventional scenario. On the other hand, their paper 
also shows that even active workers in the field may fail to grasp fully
some of the mathematical issues involved. Therefore we feel
compelled to once again attempt to clarify where the troubles lie. In
spite of the fact that our view of these matters differs from theirs,
we appreciate their efforts to elucidate these important
issues and deplore the lack of interest manifested by most particle and 
condensed matter physicists.

We begin by recalling that it is generally claimed, and repeated in the 
opening paragraph of  \cite{nnw}, that the reason for the necessity of
a non-perturbative
definition of QCD is to study its non-perturbative properties, such as 
its spectrum. As we have been stating repeatedly \cite{ps}, this is false: 
one needs a non-pertubative definition of a Quantum Field Theory because 
perturbation theory (PT) produces answers in the form of divergent 
(non-convergent) series. To interprete such series, and associate a
numerical value with them, one needs a non-perturbative definition of the 
theory. 

For theories like QCD and the two dimensional (2D) non-linear $\sigma$
models the lattice version provides the needed non-perturbative 
framework. Many interesting questions, such as the spectrum, the 
relevance of PT, etc, can be asked and have well-defined, albeit sometimes 
unknown, answers. In particular, it has been assumed for years that if a 
PT computation is free of infrared (IR) divergences, then it must be 
`right'. In the non-perturbative lattice framework, `right' can be given 
a precise mathematical meaning: the difference between the 
non-perturbative (true) answer and the PT answer truncated at given order 
must be appropriately bounded:
\bee
   |C(x,y;\beta,L)-
\sum_{i=1}^k c_i(x,y;L)\beta^{-k}|=R_k(x,y;\beta,L)=o(\beta^{-k})
\ee
Here $C$ is some Green's function, say 
$\langle s(x)\cdot s(y)\rangle$, $x$ and $y$ lattice
coordinates, ${c_i(x,y;L)}$ the PT coefficients for $C$,
$\beta$ the inverse (bare) coupling, $L$ the linear size of the lattice 
and $R$ the remainder. 
The mathematical statement that PT is providing the correct asymptotic 
expansion of $C$ in powers of $1/\beta$ is nothing but a shorthand
for the inequality (1). In many articles, conference presentations, etc 
one encounters the following meaningless statement: 'the PT series is 
asymptotic`. What is meant, is that the series is divergent. To say
that the series represents an asymptotic expansion makes sense
{\it only} if a non-perturbative definition exists and inequality (1) 
can be verified.

For $L$ fixed it is straightforward to prove (1). The subtle question is
what happens when $L$ goes to $\infty$? In particular, in order to prove
that taking the termwise limit $L\to \infty$ in eq. (1) produces the
correct asymptotic expansion of $C$ one must control the remainder R,
rather then merely prove that the limit $L\to \infty$ of $c_i(x,L)$
exists, as has been assumed for years in particle and condensed matter 
physics.   
In spite of vigorous attempts by mathematical physicists, this feat has 
been achieved so far only for Abelian cases \cite {bric}, but Niedermayer
et al claim to have found a new line of attack, which supposedly,
if not rigorous, makes it entirely plausible that the same is true for 
the non-Abelian cases. Unfortunately, as we will argue next, we find that
some of their arguments are mathematically imprecise and others
outright incorrect.

Firstly, please note that there are 
two entirely different ways to let $L$ go to $\infty$: i) independent of
$\beta$ and ii) as a given function of $\beta$. It could happen that the
remainder $R_k(x,y;,\beta,L)$ is approriately bounded in the latter case, but
not in the former. In their paper Niedermayer et al ignore 
this obviuos difference and essentially argue that they can control 
$R_k(x,y;\beta,\infty)$ by controlling $R_k(x,y;\beta,L(\beta))$ for an
appropriately chosen $L(\beta)$, namely $L(\beta)\propto\beta^k$. Such a
procedure is totally wrong: if PT is OK, then inequality (2) has to be 
obeyed at all orders for a given, ab initio made, choice of $L(\beta)$; in
other words, one cannot keep changing the function $L(\beta)$ as one tests
higher and higher orders of PT. Now if one makes the choice they 
recommend, namely $L(\beta)\propto\beta^k$ for some given k and if the PT
coefficients $c_i(x,y;L)$ admit expansions in $1/L$ (as they assume),
then their choice for 
$L(\beta)$ will change the whole expansion of $C$ in powers of $1/\beta$ (at
sufficiently high order). In fact, for their choice of 
$L(\beta)\propto\beta^k$
the PT expansion will depend upon the boundary conditions (b.c.) used, 
while the asymptotic expansion of $C(x,y;\beta,\infty)$ is clearly
independent of the b.c. used to reach the limit $L\to \infty$. Moreover
if the finite volume corrections contained terms of the form $\ln (L)/L^a$
with $a$ some integer,
as Niedermayer et al suggest, then their choice of $L(\beta)$ would 
produce an asymptotic expansion containing also $\ln(\beta)$, an
outcome which presumably they would not like for the infinite 
volume PT expansion.

A second point about which Niedermayer et al are imprecise concerns the 
distance $|x-y|$ at which $C$ is to be evaluated. Again, one could consider
many cases, and the outcome could be vastly different. For instance one 
could consider the case $|x-y|$ fixed (in lattice units), $x$ and $y$ at the
center of the lattice, $L$ going to $\infty$. This was the case analyzed by
us analytically in \cite{si}. It is not relevant for the continuum limit, 
hence our disagrement \cite{rep}
 with David's criticism \cite{dav}. For taking the 
latter limit, one must also let $|x-y|$ diverge as a given function of $L$ or
$\beta$ (see \cite{pscont}. How PT would fare in these various
possibilities is an open question. It is however clear that if it fails 
at fixed lattice distance $|x-y|$, it will also fail for limits
relevant for taking the continuum limit. Instead of stating 
precisely which limit they consider, Niedermayer et al make the 
vague statement (above their question Q1) that `all arguments are
contained in a region whose size is negligible compared to the
correlation length $\xi(\beta)$'. They should state clearly if the
distance $|x-y|$ is fixed, or else, which function of $L$ and/or $\beta$
they consider.

Next let us consider the main strategy of Niedermayer et al for controlling 
the remainder $R$. It consists in assuming that correlation inequalities,
proven for the Abelian case \cite{gin}, also apply to the non-Abelian 
models. We have no objection to this hypothesis, which leads to their eq. 
(3.1):
\bee
C^{free}(x,y;\beta,\Lambda)\leq C^\infty(x,y;\beta)\leq
C^{Dir}(x,y;\beta,\Lambda)
\ee
Essentially the whole argument of Niedermayer et al boils dowwn to the
fact that if the PT coefficients $c_r^{free}(L)$ and $c_r^{Dir}(L)$ converge
to the same values for $L\to\infty$, then, because of inequality (2), so
must the coefficients of $C^\infty(x,y,\beta)$. This is a priori false, as
can be seen from the following counterexample: suppose that the true 
functions $C^{free}(0,1;\beta,L)$ and $C^{Dir}(0,1;\beta,L)$ were given by the
following expressions:
\bee
      C^{free}(0,1;\beta,L)=1-1/\beta(1+1/L)
\ee
\bee
      C^{Dir}(0,1;\beta,L)=1-1/\beta(1-1/L)+1/\beta^2\exp(-\beta/L)
\ee
It can be easily verified that for any $\beta$ and $L$ sufficiently large, the
following is true:

\noindent
-- ${C^{Dir}(0,1;\beta,L)>C^{free}(0,1;\beta,L)}$\\
-- $C^{Dir}(0,1;\beta,L)$ is increasing in $\beta$ and decreasing in $L$\\
-- $C^{free}(0,1;\beta,L)$ is increasing in both $\beta$ and $L$ \\
-- For $L(\beta)\propto\beta$, the remainder $R_1^{Dir}(\beta,L(\beta))$ obeys
their eq. (3.7b).

Consequently these expressions obey the conjectures made by 
Niedermayer et al. Nevertheless
\bee
      C^{free}(0,1;\beta,\infty)=1-1/\beta
\ee
while
\bee
      C^{Dir}(0,1;\beta,\infty)=1-1/\beta+1/\beta^2
\ee
Consequently, even if we accept their answer to their question Q2, 
the validity of correlation inequalities for non-Abelian models and their 
conjecture for the remainder eq. (3.7b), contrary to their claim, 
the answer to their question Q3 could be a resounding NO. Indeed, having 
ruled out the second part of their question Q3, let us add that the 
counterexample given above could easily be modified by taking the additional 
(non-perturbative) term to be of the form 
$f(\beta)\exp(-\beta/\L)$ with
$f(\beta)$ a decreasing function of $\beta$ which does not possess an
asymptotic expansion in $1/\beta$ (such as $\ln(\beta)/\beta^3$).

Before abandoning the discussion of this part of the paper by
Niedermayer et al,
let us point out two other clear errors related to their
`heuristic proof that PT is OK':\\
1) Below eq. (3.9) they state incorrectly that David proved the 
cancellation of IR divergences for the {\it remainder} of PT. This is 
false and in fact, such a feat would essentially amount to a
proof that PT is correct. In fact, David's proof for the cancellation of 
IR divergences in $O(N)$ invariant Green's functions was performed in the
continuum, using a magnetic field IR regulator and dimensional 
regularization. We are not aware of any similar proof for a lattice 
model with any type of b.c.. There is also no proof that for 
${\L\to\infty}$ free and Dirichlet b.c. give the same PT expansion, as they
assume in the answer to their question Q2 (the statement may have been 
verified up to $O(1/\beta^3$ at best). \\
2) Their `heuristic argument' in favor of eq. (3.9), is incomprehensible.
While we see no connection between
$R_k^\alpha(\beta,L)$ in (3.9) and $\bar R(\beta,\delta)$ in (3.12), even
according to their own estimate the bound in eq. (3.14) does not look 
as that in eq.(3.9).

In conclusion, the claim of Niedermayer et al, that they have a 
heuristic proof that taking the termwise limit ${L\to\infty}$ of PT with
free and/or Dirichlet b.c. produces the correct infinite volume expansion 
of fixed lattice distance $O(N)$ invariant Green's functions, is 
unfounded. One
could introduce an additional assumption and provide a rigorous proof. We 
will sketch the argument, even though we see no justification for these 
assumptions. Suppose that besides their assumption that PT expansions with 
free and Dirichlet b.c. converge to the same termwise limit for 
$L\to\infty$ and that the convergence is like $1/L$ (up to $\log(L)$ powers)
and that correlation inequalities are valid, 
we asssume that the derivative of $C^\alpha(x,y;\beta,L)$ 
with respect to $L$
vanishes less fast than $1/L^2$ for ${L\to\infty}$ (as an example, please
note that this would happen 
if in our counterexample we replaced $\exp(-\beta/L)$ by
$\exp(-\beta/\ln(L))$.
Now our counterexample does not work anymore because now, 
for $\beta$ large but fixed, for $L$ sufficiently large 
$C^{Dir}(x,y;\beta,L)$ is
increasing in $L$, in violation of the conjectured correlation
inequalities. Therefore, if one could prove that: \\
1) Free and Dirichlet b.c. converge to the same limit as $1/L$ \\
2) The remainder is bounded by $1/\beta^k\exp(-\beta/\ln(L))$, \\
then either the expansion of $C(x,y;\beta,\infty$) would be given by the
(naive) PT or correlation inequalities would not hold for the model. For 
the $O(2)$ model, presumably these assumptions are correct, however, as 
we discuss below, the IR divergences are expected to be milder.

Next let us discuss their claims regarding super-instantons (s.i.). 
Firstly, below their eq. (2.32) they claim incorrectly that in our paper 
\cite{si} we stated that PT with s.i. b.c. produces IR finite answers. 
This is false and we never claimed anything like that:
we showed only that at order $1/\beta^2$ the answers
are IR finite, yet different from those obtained with periodic b.c. {\it 
only} for non-Abelian models. The latter point, regarding this manifest 
difference between Abelian and non-Abelian models, which we both verified 
and explained in our paper \cite{si}, is totally ignored by Niedermayer 
et al. In fact, if s.i. b.c. are `sick', as they and David \cite{dav},
would like to argue, how come they are alright for the $O(2)$ model? Or are
Niedermayer et al claiming that even for $O(2)$, the IR divergence they
claim to have found at $O(1/\beta^3)$ is present? This is an important
point which they (and David) should address. It was similarly ignored by 
Brezin, David and Zinn-Justin (ref.20 in the Niedermayer et al paper) 
when they tried to argue that in $1D$ the IR divergences occur 'for
dimensional reasons`: this is clearly false, since dimensional analysis 
works the same way for $O(2)$ and for $O(N)$ $N>2$ models. Our explanation
(see \cite{si}) for this difference is that only for $O(2)$ is the Gibbs
measure a function of gradients, hence IR finite (to see this, 
parametrize the spin as $(\cos(\phi(x),\sin(\phi(x)))$.

Secondly, assuming that indeed PT with s.i. b.c. does become IR divergent 
at sufficiently large order, while with say periodic b.c. not, does it 
mean that taking the termwise limit ${L\to\infty}$ of the latter produces
the correct infinite volume expansion? The mathematical answer is clearly 
NO, since what is important for asymptoticity is control of the 
remainder, not merely finiteness of the terms. It should finally be
remarked that it is even a stronger failure of the perturbative method
if different b.c. not ony give different results, but some give finite
and others infinite answers. Since it is a priori not clear that the true
infinite volume expectations actually have asymptotic expansions in
invers powers of $\beta$, it is even conceivable that an infinite answer
is correct in the sense that it shows the failure of such an expansion.

But the reason to doubt PT is more 
serious than the mere absence of a mathematical proof. What we have 
stressed over the years \cite{pat} \cite{f&f} \cite{sg}, is that PT is a 
saddle point expansion 
and for such a procedure to work, two conditions should be met: \\
1) the saddle should be `sharp', \\
2) the saddle should be far from the edge of the integration region.\\
In $O(N)$ models, on an infinite lattice, the Mermin-Wagner theorem
guarantees that the saddle cannot be sharp. While this has been known 
for years, in our papers \cite{si} \cite{sg} we showed that in the
infinite volume limit super-instanton 
configurations become degnerate with the trivial vacuum; consequently
{\it any} correct
PT expansion, irrespective of the b.c. used, must include their 
contribution for $L\to\infty$. 

So if in fact Niedermayer et al are
right that PT with s.i. b.c. becomes IR divergent at $O(1/\beta^3)$, so
does the correct infinite volume PT, whether the infinite volume is
reached via free or Dirichlet or any other legitimate b.c.,
since a correct saddle point
expansion should include expansions around many super-instanton 
configurations. From the double well harmonic osillator is has been
learned long ago that it is crucial to include such configurations
that are nearly degenerate with the ground state (in that case a
gas of instantons and anti-instantons) in order to reproduce the
correct asymptotics in the semiclassical limit (see for instance 
\cite{gp}). From the experience with that model one might expect
that the effects of the super-instantons gas might be reproduced
if one includes {\it all} saddle points, including the ones in the
complexified spin space \cite{rr}.
Niedermayer et al fail to appreciate this point. In their attempt
to prove that s.i. b.c. are `sick' they make another slip:
namely their eq. (4.4) is wrong, the correct equation
being:
\bee
\langle {\cal O}\rangle_{Dir}= {\int d{\bf S}_{z_o}
\langle {\cal O}\rangle_{S_{z_o}} e^{-\beta F({\bf S}_{z_o})}
\over \int d{\bf S}_{z_o} e^{-\beta F({\bf S}_{z_o})}}.
\ee
But even more surprising is that they don't seem to notice that their
arguments for the `sickness' of s.i.b.c. would equally well apply to
the $O(2)$ model, where in fact there is no difference between Dirichlet
and s.i.b.c..

Before concluding, let us make another point regarding finiteness versus 
correctness of PT: in $1D$ it is true that free b.c., which give a finite
answer, give the correct answer. 
The reason is that in $1D$ one knows the highest eigenvector of the 
transfer matrix (`ground state'), which is just a constant on the sphere,
and free b.c. only project onto that eigenvector, making the
expectation values independent of $L$.
No such simple L
dependence occurs in $2D$ with any b.c., hence there is no reason to make
any analogy between finiteness and correctness with the $1D$ case.

In the final paragraphs of their paper Niedermayer et al reiterate the 
standard non-perturbative arguments in favor of the standard dogma. We 
have answered many times these arguments, which we find wanting 
\cite{f&f}. Let us briefly recap:\\
- In \cite{1/n} we showed that in the $1/N$ expansion the limits
${N\to\infty}$ and ${\beta\to\infty}$ do not necessarily commute for
${L=\infty}$.\\
- For $O(3)$ the Bethe ansatz prediction for $m/\Lambda$ \cite{hn} is larger
than its MC value by about $15\%$. Of course, if our prediction that there
is a transition to a massless phase at finite $\beta$ is correct, then at
some $\beta$ the MC value for $m/\Lambda$ must cross the predicted value,
however with a non-vanishing slope.\\
- The MC data (produced by us) testing the bootstrap S-matrix prediction
are not sufficiently well under control to say whether that prediction is 
correct (we find large lattice artefacts). This is the situation at
low $p/m$. At large $p/m$ both the lattice artefacts and the S-matrix
prediction are under even poorer control, so to claim, as Niedermayer et 
al do, that the results coincide with renormalized PT and show asymptotic 
freedom is a gross exaggeration. In fact the MC data suggest that while
the S-matrix prediction could be right, it most likely {\it disagrees 
with asymptotic freedom}: indeed, we find \cite{psdod} that MC data for the
dodecahedron model are indistingushable from those for $O(3)$, but the
former can surely not possess asymptotic freedom (since at sufficiently 
large $\beta$ it possesses long range order).

While we find these arguments in favor of the accepted dogma wanting, we 
believe that our percolation arguments in favor of the existence of a 
masless phase in {\it all} $O(N)$ models are much more compelling and
under better theoretical control. Indeed in \cite{alg} and \cite{jsp}
we proved rigourously that for a different version of the $O(N)$ models,
the so called cut-action in which the spin gradient is restricted, either 
a certain well defined `equatorial cluster' percolates or the model must be
massless. Although five years have passed, no mathematical physicist has 
provided us, either in print or in private, with any heuristic arguments 
of how this equatorial cluster could possibly percolate. Moreover, as we 
stated in those papers \cite{alg} \cite{jsp}, if the equatorial cluster does 
not percolate, the typical configuration must be such that the inverse
image of 
any sufficiently large piece of the sphere forms clusters of arbitrarily 
large size.  As we emphasized in \cite{si}, such scale invariant
configurations which we believe should be the typical configurations at
low temperatures are very much like a gas
of super-instantons, an independent observation, which came 3 
years after the percolation arguments were written down.
It appears to us that to close one's eyes to these facts, as 
Niedermayer et al do, and simply reiterate 
the standard lore is not very scientific. If they find anything
wrong with our percolation arguments, they should explain it; if not, 
they should worry that the standard picture may after all be wrong.

}                                                                               

\end{document}